\documentstyle[epsf,times,raulbib]{mn}
\title{The r\^{o}le of star formation in the Tully-Fisher law} 
\author[A.F. Heavens \& R. Jimenez]{A.F. Heavens \& R. Jimenez\\ 
Institute for Astronomy, University of Edinburgh, 
Blackford Hill, Edinburgh EH9 3HJ, U.K.}

\newcommand{\be}{\begin{equation}}
\newcommand{\ee}{\end{equation}}
\newcommand{\ba}{\begin{eqnarray}}
\newcommand{\ea}{\end{eqnarray}}

\begin{document}
\maketitle

\begin{abstract}
We investigate the influence of the star formation rate on the
Tully-Fisher relation.  We find that a simple model which combines the
empirically-determined star-formation rate with the expected
properties of galaxy halos provides a remarkably good fit to the
absolute magnitude-rotation speed correlation.  We find that the
power-law nature, its slope, normalisation and scatter, are all
readily accounted for if the Universe has a low density parameter,
with or without a cosmological constant and disks are assembled at $z
\sim 1 - 1.5$. Moreover, this agreement is found simultaneously in 4
wavebands.  An Einstein-de Sitter Universe produces disks which are
too faint unless the disks are assembled at $z\sim 0.5$.  The scatter
in the relation is due to a combination of the expected range of spin
parameters of the halos and the range of formation redshifts. The
source of the scatter opens up possibilities of a better galaxy
distance indicator, if spectroscopic observations of globular clusters
can be used to determine the halo rotation.
\end{abstract}

\section{Introduction}

The correlation between the luminosity and the rotation speed of
spiral galaxies ($L \propto v^\alpha$, with $\alpha \sim 3$) has been
used as a distance indicator for nearby galaxies (see
e.g. \pcite{SW95}) since its discovery twenty years ago \cite{TF77}.
In the very roughest terms, the correlation is explicable as a
consequence of the virial theorem, but the fact that the slope of the
relation is dependent on the waveband used suggests that the picture
is at best slightly more complicated.  There are clearly at least two
potentially relevant ingredients in a model to explain the
Tully-Fisher relation, these being the properties of the mass
distribution (the halo) and the stellar properties of the baryonic
disk material.  \scite{MMW98} assume the disk mass-to-light ratio is
constant, and therefore argue that it is the halo properties which
control the Tully-Fisher relation.  On the other hand \scite{Silk97}
argues that self-regulated star formation is the controlling
influence, and the Tully-Fisher relation has nothing to do with the
halo.  The picture may be complicated still further by effects such as
energy feedback from stars, and fresh infall of gas over time,
motivating an approach where all physical processes are modelled with
complex computer codes (e.g. \pcite{SN98}).  In this paper, we take a
simpler complementary approach, which combines halo properties with
empirical star formation properties.  This approach has the advantage
that it offers a possibility of improving the correlation by measuring
further physical parameters; this is the main motivation for taking
this analytic approach.  It also appears to account for the relation
extremely well.  We model the halo with the theoretically known spin
properties of collapsed objects, using the isothermal sphere for the
density profile, partly for simplicity, and partly because the issue of 
the dark matter profile in collapsed haloes is not yet settled 
\cite{Moore98}.  

We model the star formation entirely
empirically, using the Schmidt law relating star formation rate to
disk surface density \cite{Kenn98}.  With this simple model, we are
able to obtain analytic expressions for the star formation rate as a
function of time.  The spectrophotometric evolution cannot be done
accurately in an analytic way, so for this step we use a synthetic
stellar population numerical code \cite{JPMH98,JDPPMJ99}.  Rather
surprisingly, this simple model is able to account extremely well for
the Tully-Fisher relation.  We are able to reproduce the slope,
normalisation and scatter of the relation, simultaneously in 4
wavebands, with very little freedom.  The idea of the spin parameter
influencing the galaxy properties is of course not new; it has been
standard since \scite{FE80}, and applications have been made for disk
galaxies for example by \scite{K82}, \scite{Dalc97}, \scite{JHHP97},
\scite{JPMH98}, as well as \scite{MMW98,B98}.  The new element in the
analysis here is the combination with the empirical dependence of star
formation rate on surface density.

\section{Halo and star-formation model}

We follow the notation and method of \scite{MMW98} in the modelling of
the halo by an isothermal sphere, characterised by its mass $M$, and
circular velocity $V_c$.  These are related via the Hubble constant at
the redshift $z$ of formation $H(z)$ \be M = {V_c^3\over 10 G H(z)}.
\ee Note that $V_c$ is not the actual circular velocity of the halo --
it is the circular velocity required for centrifugal support in the
potential of the halo.  In terms of the present Hubble constant and
the density parameter in non-relativistic matter and cosmological
constant, $H(z)=H_0\left[\Omega_{\Lambda 0} + (1- \Omega_{\Lambda
0}-\Omega_0)(1+z)^2+\Omega_0(1+z)^3\right]^{1/2}$.  The disk mass we
assume to be $M_d = m_d M$, where the fraction $m_d=\Omega_b/\Omega_0$
and the baryon density is set by nucleosynthesis \cite{Walker91},
$\Omega_b = 0.015 h^{-2}$, where $h\equiv H_0/100$ km s$^{-1}$
Mpc$^{-1}$ is taken to be 0.65 throughout.

We assume the disk settles with an exponential profile
$\Sigma(R)=\Sigma_0 \exp(-R/R_d)$, with central surface density
$\Sigma_0$ and scale length $R_d$, in which case $M_d = 2\pi \Sigma_0
R_d^2$.  The disk scale length is related to the spin parameter
$\lambda$ of the halo by \cite{MMW98} \be R_d = {\lambda V_c\over 10
\sqrt{2} H(z)} \ee where we have assumed that the disk and halo have
the same specific angular momentum \cite{Mestel63}.
$\lambda=J|E|^{1/2}G^{-1}M^{-5/2}$, where $E$ is the total energy of
the halo, and $J$ its angular momentum.  This sets the initial gas
surface density, which we assume is accreted on a timescale which is
short compared with a Hubble time.  We also ignore any gas returned to
the ISM by stars, or late infall of fresh gas.

We assume that the star formation rate is set by the empirical Schmidt
law \cite{Kenn98}, dependent only on the local gas surface density
$\Sigma_g$, $\Psi_{\rm SFR} = B {\Sigma_g}^{1.4 \pm 0.15}$: 
\be 
\Psi_{\rm SFR} = (2.5\pm 0.7) \times 10^{-4}
\left({\Sigma_g\over M_\odot pc^{-2}}\right)^{1.4\pm 0.15} M_\odot
y^{-1} kpc^{-2}.  
\label{Sch}
\ee 
The time-evolution of the gas surface density is
therefore given in terms of the initial gas surface density
$\Sigma_{g0}$ by 
\be 
\Sigma_g(t) = \left(\Sigma_{g0}^{-0.4}+0.4
Bt\right)^{-2.5} 
\ee 
where $B=9.5 \times 10^{-17}$ in SI units and $t$
is measured from the assembly of the disk. From this and the star
formation law, we can integrate over the disk to compute the total
star formation rate: 
\be 
\dot{M}_*(t) = {2\pi B R_d^2
\Sigma_0^{1.4}\over (0.4)^2}\int_0^\infty \,dy\, y\left(\exp
y+a\right)^{-3.5} 
\ee 
(in kg s$^{-1}$), where $a(t,\Sigma_0) \equiv
0.4 Bt\Sigma_0^{0.4}$.  This may be written in terms of a generalised
hypergeometric function (e.g. \pcite{GradRyz}): 
\be 
\dot{M}_*(t) =
{50\pi B R_d^2 \Sigma_0^{1.4}\over 49} \phantom{I}_3
F_2\left(3.5,3.5,3.5;4.5,4.5;-a\right).  
\ee 
A similar integral gives
the remaining gas mass as a function of time as 
\be 
M_g(t) = {2\pi
R_d^2 \Sigma_0} \phantom{I}_3
F_2\left(2.5,2.5,2.5;3.5,3.5;-a\right).  
\ee 
In more astronomical units:
\begin{eqnarray}
\dot{M}_*(t) & = &  247 h_z^{-0.6}\left({m_d\over 0.05}\right)^{1.4} 
\left({V_c\over 250 km s^{-1}}\right)^{3.4} 
\left({\lambda\over 0.05}\right)^{-0.8}  \times \nonumber\\
& & \phantom{I}_3 F_2\left(3.5,3.5,3.5;4.5,4.5;-a\right)\ M_\odot
y^{-1}
\end{eqnarray}
and 
\be 
a = 1.06 h_z^{0.4}\left({t\over Gyr}\right) \left({V_c\over
250 km s^{-1}}\right)^{0.4}\left({\lambda\over
0.05}\right)^{-0.8}\left({m_d\over 0.05}\right)^{0.4}.  
\ee 
$h_z \equiv H(z)/100$ km s$^{-1}$ Mpc$^{-1}$.  We therefore have a
complete description for the star formation rate as a function of time, for
any cosmology and any halo parameters $V_c$, $z$ and $\lambda$.  The 
relative star formation rate and remaining gas fraction are shown 
in Fig. 1. The gas fractions remaining depend sensitively on cosmology and 
circular velocity but are typically in the range $1 - 10\%$ 
(c.f. \scite{RH94}).
\begin{figure}
\begin{center}
\setlength{\unitlength}{1mm}
\begin{picture}(90,60)
\includegraphics{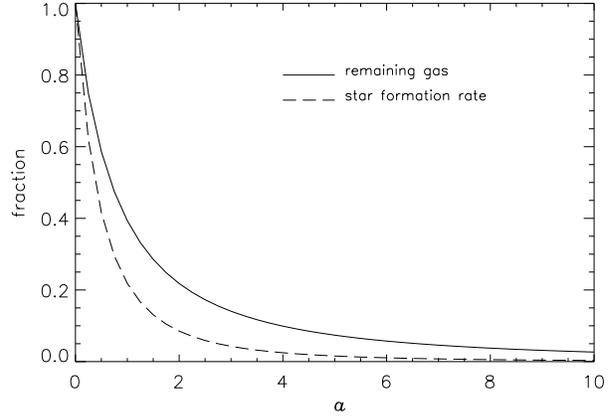}
\label{SFR}
\end{picture}
\end{center}
\caption{The star formation rate and remaining gas fraction, normalised to 
unity at the disk formation time, as a function of the parameter $a \propto 
t$.}
\end{figure}
The star formation rate is then fed into a sophisticated
spectrophotometric stellar evolution code, which has been described
elsewhere \cite{JPMH98,JDPPMJ99}.  Broad-band magnitudes in $B$, $R$,
$I$ and $K'$ are obtained from the resulting spectra.  Note we have
chosen a particular synthetic population code; other codes, such as
those discussed in \scite{CWB96}, agree to within 0.2 magnitudes.
This leads to a small shift in normalisation, but does not add
dispersion.  We assume a solar metallicity throughout.  Including
chemical evolution would lead to higher spin systems being bluer, and
lower spin systems being redder than we have computed, but the effect
is confined to B.

\section{Results}

We show the results as graphs of absolute magnitude against circular
velocity, in $B$, $R$, $I$ and $K'$, for disk formation redshifts
$z$=1,2,3 and spin parameters $\lambda$=0.025, 0.05, 0.1.  These spin
parameters represent the 10, 50 and 90 percentile points for a
gaussian field for any reasonable power spectrum
(e.g. \pcite{Warren92}).  Disks are subject to disruption by major
merging events, and there are arguments that present-day disks may
have been assembled at $z \sim 1-2$ \cite{WEE98}, and there are
arguments from disk sizes that argue for even lower assembly
redshifts \cite{MMW98}.  In Fig. 2, we show
the results for a flat cosmological model with $h=0.65$,
$\Omega_0=0.3$, $\Omega_{\Lambda 0}=0.7$.  Superimposed is the data
from the careful study of spiral galaxies in the loose clusters of
Ursa Major and Pisces \cite{Tully98}.  The inclination-corrected FWHM
of the lines has been converted there to $W_R^i$, which approximates to 
twice the circular velocity. From Figure 2 and also for the 
model of Figure 4, we find the following:
\begin{figure}
\begin{center}
\setlength{\unitlength}{1mm}
\begin{picture}(90,120)
\includegraphics{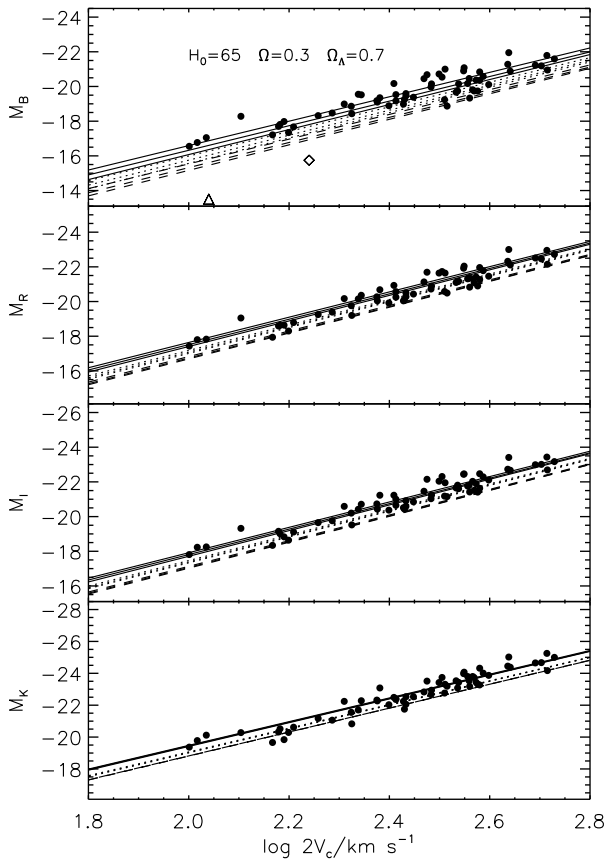}
\label{flat}
\end{picture}
\end{center}
\caption{The absolute magnitude (in 4 bands) vs. circular velocity for 
disk galaxies in a flat, $\Lambda$-dominated universe, with
$\Omega_0=0.3$, $\Omega_{\Lambda 0}=0.7$, and $H_0=65$ km s$^{-1}$
Mpc$^{-1}$.  The solid, dotted and dashed lines correspond to disk
formation redshifts of 1,1.5 and 2 respectively.  For each formation
redshift, the 3 lines show halos with spin parameters $\lambda=0.025$,
0.05 and 0.1 (bottom-to-top).  These spin parameters are the 10, 50
and 90 percentile points for the spin distribution from gaussian
initial conditions.  The data are drawn from Ursa Major and Pisces
clusters (Tully et al. 1998).  Also plotted are the near-dark galaxies
NGC2915 (diamond) and DDO154 (triangle).}
\end{figure}

\begin{itemize}
\item{The T-F relation is a good power-law, with about the right slope}
\item{The T-F normalisation is reproduced for plausible formation redshifts}
\item{The scatter is comparable to observation, if disk formation redshifts 
are not too widely distributed}
\end{itemize}
Moreover, these successes apply simultaneously to all four colours
(especially for $R$, $I$ and $K'$, which sample the old population).
The blue data are a little more ragged; nevertheless our model
provides a good fit leaving little room for any enhanced star
formation due to late infall of gas. It is worth noting that the
predicted level of star formation agrees well with that found by
\scite{M98,S99} since our model has typical initial SFR of 10-100
M$_{\odot}$ y$^{-1}$ with a highest possible value of 1000
M$_{\odot}$ y$^{-1}$ for the systems with the highest $V_c$ and
lowest $\lambda$.

The magnitude is not necessarily a monotonic function of spin
parameter, at fixed $V_c$.  Low-spin systems forming at given $z$ will
have high surface densities and high initial star formation rates.
At first they will be brighter than the high-spin systems.  However,
the timescale over which they use up most of their gas is short, and
the fading of stars over time may make them fainter than the high-spin
systems at late times.

Fig. 3 shows results for an Einstein-de Sitter cosmology.  This model
forms disks with very high surface densities, which have high initial
star formation rates, but which fade excessively unless their
formation redshift is low ($\sim 0.5$).  One could argue that we have
a small amount of freedom in moving the curves left and right, since
the observed relation uses the FWHM $W$ of the lines, and we have
assumed $W=2V_c$.  This is probably most appropriate for the
isothermal sphere model we assume, but one might allow some
flexibility here.  We have also investigated an open universe with
zero cosmological constant; it fits almost as well as the $\Lambda$
model.

In Fig. 4, we show the flat model, but we incorporate the peak
height-spin parameter anticorrelation (approximately $\lambda \propto
\nu^{-1.4}$) claimed from analytical studies by \scite{HP88} (see also
\pcite{SB95}).  There is some support for this from the numerical
simulations of \scite{Ueda94}.  \scite{Lemson99} see no
anticorrelation with density, but they present results only for
density in a rather large sphere (10 $h^{-1}$ Mpc).  Since the density
on galaxy scales (filter length $\sim 0.5 h^{-1}$ Mpc) is only weakly
correlated with density on such large scales, it is hard to see how
any small-scale correlation could be apparent in such a study.
Assuming that the small-scale anticorrelation is present 
makes little difference to the Tully-Fisher relation, for a reasonable
mean formation redshift, which we take to be 1.5 
(Note that this refers to the mean collapse redshift of the
haloes; the disks observable today may be younger and have a lower
assembly redshift).  Early-forming high peaks have systematically
lower spin, which leads to higher surface density, brighter disks.
The predicted scatter in the relation depends on the probability
distribution of the disk assembly redshift.  We can estimate a lower
limit to the scatter by assuming a single formation redshift.  For
$z=1-1.5$, the scatter is approximately 0.3 magnitudes in $K$ and 0.7 in
$B$; the observed values in Ursa Major and Pisces vary from 0.42 in
$R$, $I$ and $K'$ to 0.55 in $B$, although a sample of 12 clusters
shows a smaller scatter of 0.35 in $I$ \cite{Tully98b}.  Inspection of
Fig. 4 shows that a range of formation redshifts can give good
agreement with the observed scatter.  Note that the uncertainty in the
Kennicutt fit to the Schmidt law (equation \ref{Sch}) makes relatively
little impact on the results; changing the normalisation of the star
formation rate by 30 per cent either way alters the magnitudes by
about 0.15.  Further we note that the anticorrelation would ease the
problem of disks being too small compared with observation
\cite{MMW98}, since the later-forming haloes have typically larger
spins than average.  Late-forming haloes forming at redshift 1 (when
the mean is 1.5) are 40\% larger if the anticorrelation holds.

The thick lines in Fig. 4 show the theoretical limit for galaxies for forming
stars according to the \scite{Toomre64} stability criterion (see also
\pcite{Kenn89}).  As discussed in \scite{JHHP97} halos with $V_{c}
\leq 40$ km s$^{-1}$ fail to form stars for any value of $\lambda$,
while for halos with $V_{c} \geq 120$ km s$^{-1}$ stars are formed in
the disk for any value of $\lambda$ as shown in Fig.~4 of
\scite{JHHP97}. The absence of faint discs is also present in other samples
(see e.g. \pcite{SW95}), although observational selection effects may be the
explanation for current data (Tully, priv. comm.).  
To illustrate this we have also plotted the
near-dark galaxies NGC2915 (diamond) and DDO154 (triangle) (see
\pcite{Meurer96}). These systems fall quite well in the scenario
described in \scite{JHHP97} since only the central parts of these
galaxies have formed stars, and have a dark HI disk. The reason why
they fall below the theoretical TF law is due to the fact that most of
its HI mass is not in stars (they have mass-to-light ratios of about
80), thus they are under-luminous.  

Fig. 5 shows the predicted Tully-Fisher relation for disks observed at
redshift 1.  If the disk assembly is also at redshift 1, then the
disks are extremely bright in all wavebands.  For disk assembly at
redshifts 1.5 and 2, the Tully-Fisher relation is generally a little
brighter, the degree of brightening depending in detail on circular
velocity, band and formation redshift.  For massive systems forming at
redshift 1.5, the systems are about $0.5-1.5$ magnitude brighter in blue,
somewhat larger than the observations of \pcite{Vogt97}, although the
samples are not large enough as yet for the situation to be clear
(\pcite{Rix97}, \pcite{Hudson98}, \pcite{Simard98}).

The model predicts some dependence of the Tully-Fisher relation on the
surface brightness of the galaxy, in the sense that higher surface
brightness disks should have higher luminosities for given rotation
speed.  The effect is not large, as witnessed by the tightness of the
predicted relations in Fig. 4, and there is also some overlap in the
surface brighnesses at fixed $M$, $V_c$ arising from a spread in disk
assembly redshifts.  Thus the coincidence of TF relations of high and
low surface brightness disks \cite{Spr95b,Zwaan95} presents no great
difficulty for the model.  More serious is the bimodal surface
brightness distribution of disks observed by \scite{Tully97} (but see
\pcite{MSM98} for an alternative view), which
does not appear naturally in this model.  The authors suggest that the
high surface brightness disks may be self-gravitating, whereas the
LSBs are dark-matter dominated.  We cannot exclude such a possibility,
as our model assumes a given halo potential.   If the passage to
self-gravitation means only more early star formation (from higher surface
densities), then the results we present are virtually unchanged: at these
wavebands, the distinction between a burst of star formation and the
ongoing trickle which our model predicts is negligible for the high
surface density disks.  Observations at $U$ would be required to
distinguish the histories.
\begin{figure}
\begin{center}
\setlength{\unitlength}{1mm}
\begin{picture}(90,120)
\includegraphics{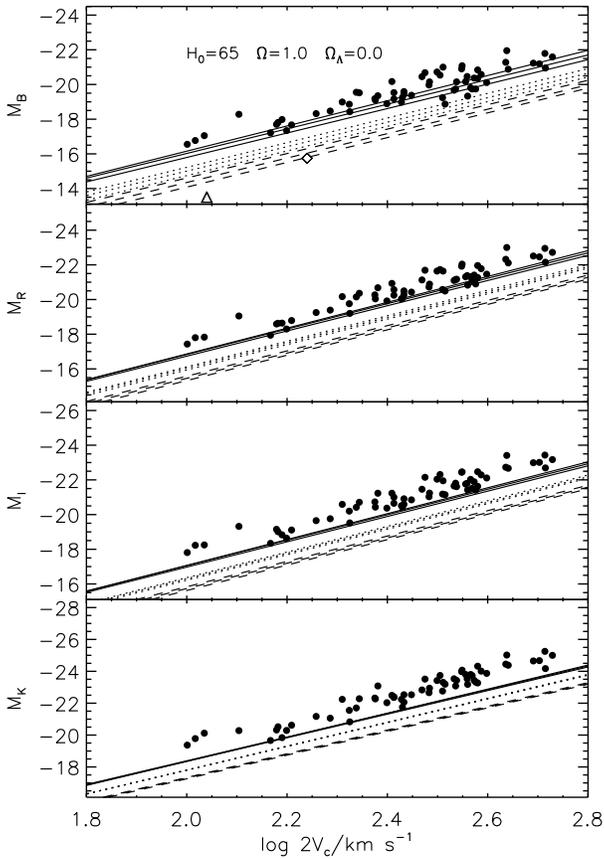}
\label{EdS}
\end{picture}
\end{center}
\caption{As Fig. 2, but for an Einstein-de Sitter universe and formation 
redshifts of 0.5 (solid line), 1.0 (dotted line) and 1.5 (dashed line).}
\end{figure}

\begin{figure}
\begin{center}
\setlength{\unitlength}{1mm}
\begin{picture}(90,120)
\includegraphics{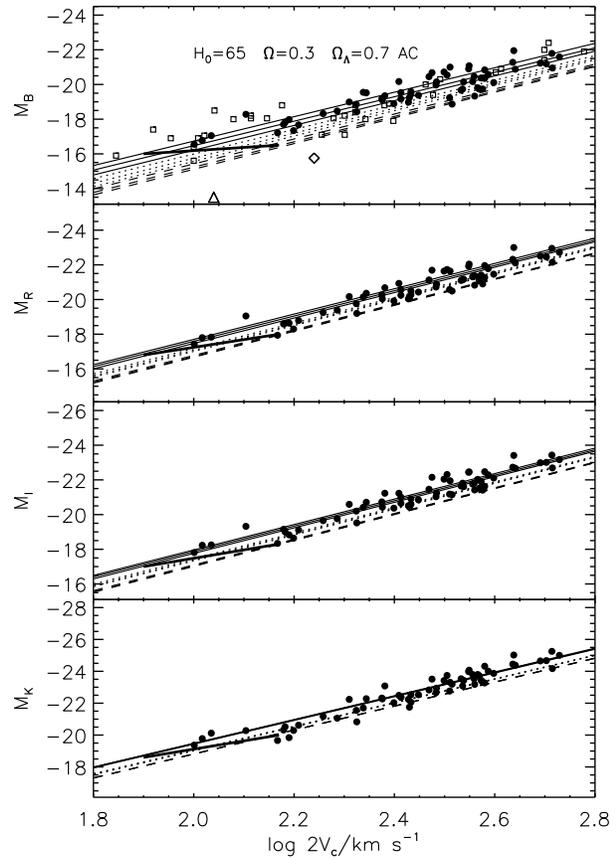}
\label{anticorr}
\end{picture}
\end{center}
\caption{As Fig. 2, but including the halo peak height-spin parameter 
anticorrelation predicted by Heavens \& Peacock (1988). The thick
solid lines shows the expected theoretical threshold for halos that
fail to form stars (Jimenez et al. 1997), for formation redshift 1.5.
System below the line will fail to form any significant population of
stars and thus will be `dark'. Comfortably enough, no galaxies are
observed below the threshold in any 4 bands, except for system with
stars confined to the central regions with few stars in the disk, such
as NGC2915 (diamond) and DDO154 (triangle).  Also plotted as open
squares are the data for LSBs from Zwaan et al. (1995), this
illustrates the point that LSB and HSB galaxies lie in the same TF law
}
\end{figure}

\begin{figure}
\begin{center}
\setlength{\unitlength}{1mm}
\begin{picture}(90,120)
\includegraphics{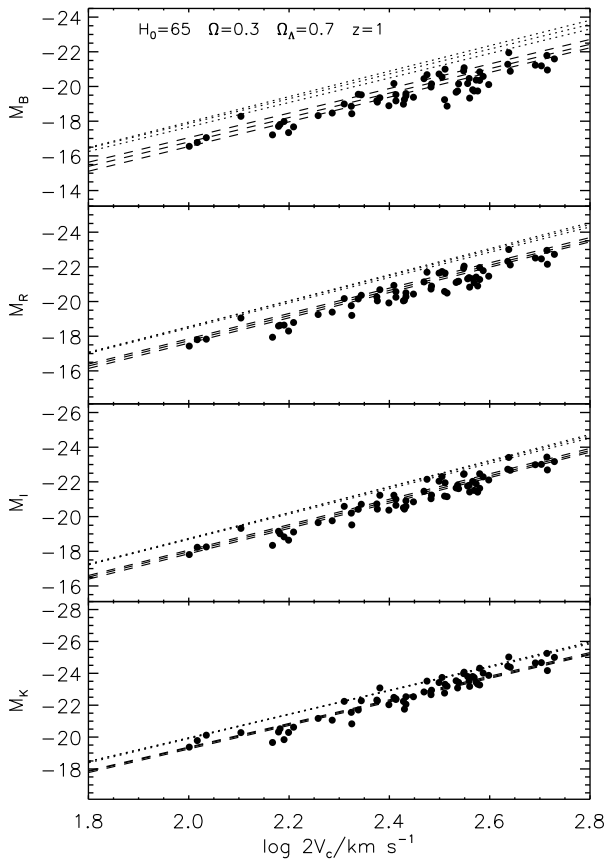}
\label{open}
\end{picture}
\end{center}
\caption{Predicted Tully-Fisher relation at $z=1$ for the $\Lambda$ 
universe of Fig. 2 and formation redshifts of 1.5 (dotted line) and 2
(dashed line).}
\end{figure}

\section{Summary and prospects for a better distance indicator}

We have combined the theoretical properties of dark matter halos with
the empirically-determined dependence of star formation rate on disk
surface density to predict a correlation between luminosity and
circular velocity which is compared with observation.  Remarkably good
agreement is found, in 4 wavebands, with the recent study of
\scite{Tully98}, for a flat universe dominated by a cosmological
constant if the disk assembly redshift is $1 - 1.5$, in agreement with
conclusions from \scite{WEE98} which are based independently on disk
survival rates.  An Einstein-de Sitter universe can be reconciled if
the disk formation redshift is low: $z\sim 0.5$.  This explanation of
the Tully-Fisher relation falls between the halo-determined extreme of
\scite{MMW98} and the halo-independent model of \scite{Silk97},
however, it is only a partial explanation, as it relies on the
empirical Schmidt law (see \pcite{Silk97} for some relevant
discussion).  This study suggests that there are two parameters which
lead to scatter in the Tully-Fisher relation.  These are the spin
parameter of the halo, and the redshift at which the disk is
accumulated.  The age of the disc appears to be less important than
the redshift, which controls the initial surface density of the disk
and hence the star formation rate.  The general effect of the
formation redshift and the spin parameter is almost inevitable, and is
likely to be more robust than the specific model calculations
presented here.  The relative influence of formation redshift and spin
parameter is model-dependent.  In particular, if, as has been claimed
\cite{HP88,Ueda94}, there is an anticorrelation between peak overdensity and
spin parameter, then the predicted Tully-Fisher relation is changed
only slightly.  The scatter might be narrowed by measurement of the
age and spin parameter.  
The latter is a difficult quantity to measure,
but could be attempted by looking at the systematic velocity of halo
stars which are not formed dissipatively in the disk.  Individual
stars would be impossible to measure, especially with a bright disk
present, but it is possible that for nearby galaxies the integrated
light from globular clusters could be used.

\section*{Acknowledgments}

We are grateful to Jim Peebles for an initial conversation which led
to this work, and particularly to Brent Tully for providing us with
the spiral sample data, and many useful comments.  We also thank Will
Saunders for helpful remarks.

\end{document}